\documentclass[aps,prl,floatfix,twocolumn,nofootinbib,superscriptaddress]{revtex4-2}
\usepackage{hyperref}
\usepackage{url}
\usepackage{setspace}
\usepackage{amsmath,amssymb,amscd,braket,amsfonts}
\usepackage{graphicx}
\usepackage{xcolor,physics,ragged2e}
\usepackage{tikz,pgfplots,xfp}
\usetikzlibrary{calc,matrix,shapes,arrows,arrows.meta,bending,decorations.pathmorphing,decorations.markings,decorations.pathreplacing,trees,positioning}
\pgfplotsset{compat=1.17}

\definecolor{groen}{RGB}{91,141,34}
\definecolor{rood}{RGB}{220,46,46}
\definecolor{paars}{RGB}{219,0,250}
\definecolor{grijs}{RGB}{204,204,196}
\definecolor{zwart}{RGB}{29,28,26}
\definecolor{roos}{RGB}{219,0,208}

\tikzstyle{singularity}=[zwart,line width=0.6,decorate,
                         decoration={zigzag,amplitude=2,segment length=6.17}]
\tikzstyle{frame}=[zwart]
\tikzstyle{blacktext}=[zwart]
\tikzstyle{uv}=[rood, line width=0.5]
\tikzstyle{r}=[groen, line width=0.5]
\tikzstyle{t}=[paars, line width=0.5]
\tikzstyle{lijn}=[roos, line width=0.5]
\tikzset{declare function={
  kruskal(\x,\c) = {\fpeval{asin( \c*sin(2*\x) )*2/pi}};
}}
\def\Nsamples{20}

\newcommand{\rme}{\mathrm e}
\newcommand{\rmd}{\mathrm d}

\newcommand{\bea}{\begin{eqnarray}}
\newcommand{\eea}{\end{eqnarray}}
\newcommand{\be}{\begin{equation}}
\newcommand{\ee}{\end{equation}}
\newcommand{\ba}{\begin{align}}
\newcommand{\ea}{\end{align}}

\begin{document}

\title{
Complexity = Anything Can Grow Forever in de Sitter
}

\author{Sergio E. Aguilar-Gutierrez}
\email{sergio.ernesto.aguilar@gmail.com}
\affiliation{Institute for Theoretical Physics, KU Leuven, 3001 Leuven, Belgium}

\author{Michal P. Heller} \email{michal.p.heller@ugent.be}
\affiliation{Department of Physics and Astronomy, Ghent University, 9000 Ghent, Belgium}

\author{Silke Van der Schueren}
\email{silkevdschueren@gmail.com}
\affiliation{Department of Physics and Astronomy, Ghent University, 9000 Ghent, Belgium}

\begin{abstract}
Recent developments in anti-de Sitter holography point towards the association of an infinite class of covariant objects, the simplest one being codimension-one extremal volumes, with quantum computational complexity in the microscopic description. One of the defining features of these gravitational complexity proposals is describing the persistent growth of black hole interior in classical gravity. It is tempting to assume that the gravitational complexity proposals apply also to gravity outside their native anti-de Sitter setting in which case they may reveal new truths about these cases with much less understood microscopics. Recent first steps in this direction in de Sitter static patch demonstrated a very different behavior from anti-de Sitter holography deemed hyperfast growth: diverging complexification rate after a finite time. We show that this feature is not a necessity and among gravitational complexity proposals there are ones, that predict linear or exponential late-time growth behaviors for complexity in de Sitter static patches persisting classically forever.

\end{abstract}

\maketitle

\section{Introduction and summary}
Understanding de Sitter (dS) space holography at a level comparable to AdS/CFT~\cite{Maldacena:1997re,Gubser:1998bc,Witten:1998qj} is an important open question in quantum gravity dating back to the early days of AdS/CFT~\cite{Strominger:2001pn,Witten:2001kn,Maldacena:2002vr}.

Key drivers of progress in AdS quantum gravity have been ideas native to quantum information theory and quantum computing, see e.g.~\cite{Harlow:2014yka,Rangamani:2016dms,Susskind:2018pmk,Almheiri:2020cfm,Chen:2021lnq,Chapman:2021jbh} for reviews. In recent years these tools have started being applied also to positively curved universes, see e.g.~\cite{Dong:2018cuv,Susskind:2021esx,Shaghoulian:2022fop,Chandrasekaran:2022cip}. The focal object in the present article is holographic complexity, which arose as a conjectured geometric counterpart of the hardness of dual state or operator preparation using limited resources on the boundary of AdS/CFT~\cite{Susskind:2014rva,Susskind:2018pmk,Chapman:2021jbh}. Considerations based on quantum circuit models of the boundary Hamiltonian time evolution led to two defining features of such geometric quantities in AdS black hole spacetimes: late-time linear growth with time and switchback effect accounting for a delay in the late-time growth due to external perturbations (shock waves). Between 2014 and 2016 three such geometric quantities were identified and thoroughly studied over the past decade: codimension-one boundary-anchored maximal volume slices (CV)~\cite{Stanford:2014jda}, gravitational action in the Wheeler-de Witt patch (CA)~\cite{Brown:2015bva} and spacetime volume of the Wheeler-de Witt patch (CV2.0)~\cite{Couch:2016exn}. 

The approach to dS holography that is relevant for our article is the stretched horizon one~\cite{Susskind:2021esx,Shaghoulian:2022fop,Susskind:2022dfz,Lin:2022rbf,Lin:2022nss,Bhattacharjee:2022ave,Susskind:2023hnj,Susskind:2023rxm}. It can be thought of to mimic AdS holography in the native to holographic complexity setting of eternal AdS Schwarzschild black holes~\cite{Maldacena:2001kr}. The exterior of the latter corresponds to two dS static patches and the AdS asymptotic boundary is mimicked by two stretched horizons, see Fig.~\ref{fig:heuristic idea}. Since holographic complexity proposals are geometric constructs, there are no fundamental obstacles to studying them also in this setting. Indeed, over the course of the past two years first CV in two spacetime dimensions~\cite{Susskind:2021esx,Chapman:2021eyy,Galante:2022nhj} and subsequently in~\cite{Jorstad:2022mls} also CV, CA and CV2.0 in quite a generality were studied in dS stretched horizon holography, see also \cite{Auzzi:2023qbm,Anegawa:2023wrk,Anegawa:2023dad,Baiguera:2023tpt} for related recent developments. The common outcome of these studies is holographic complexity diverging (its time derivative diverging in two-dimensional dS) after a finite stretched horizon time. This hyperfast growth~\cite{Susskind:2021esx} is in stark contrast with predictions of holographic complexity proposals for AdS black holes exhibiting a persistent linear growth at a classical level consistent with a local quantum circuit model and might signal a very nonlocal nature of stretched horizon degrees of freedom.

In parallel to the first works studying holographic complexity in dS, it was realized that the space of holographic complexity proposals contains infinitely many members~\cite{Belin:2021bga,Belin:2022xmt}. Such Complexity = Anything proposals (CAny) are defined by obeying the late-time linear growth and switchback effect for AdS black holes and can be defined by codimension-1 as well as codimension-0 geometric objects. However, a priori it is not guaranteed that their other behaviors, in particular in dS, will also be shared with CV, CA, and CV2.0. This leads to our motivating question: 
\begin{quote}
Assuming stretched horizon holography and holographic complexity proposals, is the hyperfast growth as universal for dS holographic complexities as linear growth and switchback for AdS ones?
\end{quote}
Contrary to expectations stemming from the accelerated expansion of dS universes,
our work demonstrates that hyperfast growth is not a necessity within CAny proposals, but a feature appearing for some of them, with different kinds of growth present for another subset. We demonstrate this using a family of CAny proposals defined on constant mean curvature (CMC) spatial slices with our arguments covering also Schwarzschild dS black holes (SdS). However, these proposals might not describe holographic complexity as it is defined in the AdS context; instead, they are general observables of interest for static patch holography in dS space. An interpretation of holographic complexity in dS space would require developing a quantum circuit interpretation of its quantum mechanical dual theory, which is still missing. The most promising case is certainly a two-dimensional one, where on one hand our findings apply and, on the other, an expanding patch of a dS geometry can be embedded outside a horizon in an AdS spacetime subject then to more standard holographic interpretations~\cite{Anninos:2017hhn,Anninos:2020cwo,Ecker:2022vkr,Chapman:2021eyy}. Our perspective is, however, to develop geometric intuitions about what is possible for holographic complexity rather than to provide a microscopic interpretation, which apart from one isolated case~\cite{Rabinovici:2023yex} has not been settled in a precise manner.

Finally, our considerations of holographic complexity in dS universes bear implications on a subclass of CAny holographic complexity proposals in AdS utlizing CMC spatial slices. To this end, we observe that complexity interpretation in eternal AdS black hole spacetimes may generically require an additional ingredient in these CAny proposals that renders their time evolution symmetric.

\section{Setup}
The asymptotically dS geometries of interest in $d+1$ spacetime dimensions are described by the metric
\begin{equation}\label{eq:static}
    \rmd s^2=-f(r)\rmd t_{L/R}^2+\frac{\rmd r^2}{f(r)} +r^2 \rmd\Omega_{d-1}^2~,
\end{equation}
where
\begin{equation}\label{eq:blackening factor}
    f({r})=1-{r}^2-\tfrac{2\mu}{{r}^{d-2}}~
\end{equation}
and $\rmd\Omega_{d-1}^2$ is the metric on a unit ($d-1$)-dimensional sphere. Meanwhile, the (dimensionless) parameter $\mu$,
\begin{equation}   
\mu\in[0,\,\mu_N],\quad\mu_N\equiv\tfrac{1}{d}\qty(\tfrac{d-2}{d})^{\tfrac{d-2}{2}}~.
\end{equation}
allows us to study spacetimes from the empty dS ($\mu = 0$) all the way to the Nariai black hole space ($\mu = \mu_{N}$), i.e. the largest black hole that can fit in dS space. Note that in this paper we set the curvature scale associated with a cosmological constant (both positive and negative) to unity. The coordinates~\eqref{eq:static} are Schwarzschild coordinates and cover the region outside the horizon (the static patch for dS), hence the presence of two-time variables, one for each exterior.

In analogy with AdS holography~\cite{Maldacena:2001kr} and following~\cite{Susskind:2021esx,Chapman:2021eyy,Galante:2022nhj,Jorstad:2022mls}, we will be interested in introducing stretched horizons at $r = r_{\rm{st}}$ with constant $t_{L}$ and $t_{R}$ slices thereof defining states in a putative microscopic description involving two Hilbert spaces, one for each stretched horizon. We orient both time directions to increase towards future infinity and consider left-right symmetric time evolution in $t_{L} = t_{R} \equiv \frac{t}{2}$. The venerable CV proposal amounts to finding stretched horizon anchored codimension-1 volumes and studying them as a function of~$t$. Since this and any other holographic complexity proposal require connecting two boundaries through an inflating region complementary to the static patch, in explicit calculations we will be using ingoing Eddington-Finkelstein (EF) coordinates given by
\begin{equation}\label{eq:starting metric}
    \rmd s^2=-f({r})\rmd {v}^2+2\rmd {v}\rmd {r} + r^2 \rmd \Omega_{d-1}^2~.
\end{equation}
Because of the left-right symmetry, it will be enough to consider only one patch of such coordinates.

\section{Key Idea}
\begin{figure}[t!]
    \centering
    \begin{tikzpicture}[scale=2.45]
        \draw[frame] (-1,-1)--(-1,1)--(1,1)--(1,-1)--cycle;
        \node[frame, anchor=north] at (0, -1){\large ${\cal I}^-$};
        \node[frame, anchor=south] at (0, 1){\large ${\cal I}^+$};
        \draw[uv] (-1,-1) -- (1,1);
        \draw[uv] (1,-1) -- (-1,1);
        \draw[orange] (-1,0.6)  -- (1,0.6);
        \draw[orange] (-1,0.7)  -- (1,0.7);
        \draw[orange] (-1,0.8)  -- (1,0.8);
        \draw[orange] (-1,0.9)  -- (1,0.9);
        \draw[lijn] (-1,0)--(0,1)--(1,0)--(0,-1)--cycle;
        \draw[lijn] (-1,0)--(1,0);
        \node[lijn] at (0.75,0.377) {\large $\tau_\infty$};
        \node[lijn] at (0.77,-0.377) {\large $-\tau_\infty$};
        \def\Nlines{3}
        \foreach \i [evaluate={\c=\i/(\Nlines+1); \cs=sin(90*\c);}] in {1,...,\Nlines}{
        \draw[lijn, samples=\Nsamples, smooth, variable=\x, domain=0:2]
            plot(1-\x,{-kruskal(\x*pi/4,\cs))});
        \draw[lijn, samples=\Nsamples, smooth, variable=\x, domain=0:2]
            plot(1-\x,{kruskal(\x*pi/4,\cs))});
        }
        \node[r, anchor=south] at (0.9, -0.75) {\large ${r}_\text{st}$};
        \draw[r, samples=\Nsamples, smooth, variable=\y, domain=0:2]
                plot({1-kruskal(\y*pi/4,sin(45))},\y-1);
        \draw[r, samples=\Nsamples, smooth, variable=\y, domain=0:2]
                plot({-1+kruskal(\y*pi/4,sin(45))},\y-1);
    \end{tikzpicture}
    \caption{Penrose diagram of dS$_{d+1}$ space, where the stretched horizon is shown in green at ${r}_{\rm st}$ and the extremal volumes of the CV proposal in pink. The origin of the hyperfast growth is approaching the infinity touching lightcone in finite stretched horizon time (i.e. at $\tau_{\infty}$ and $-\tau_{\infty}$). In orange, we display slices of constant global time, which exhibit persistent growth as they avoid future infinity. The key idea of our paper is to find analogous slices, but belonging to CAny and understand their properties.}
    \label{fig:heuristic idea}
\end{figure}
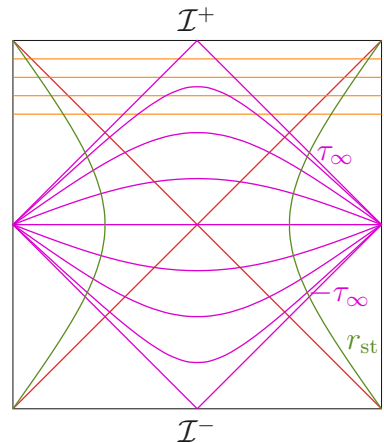

Fig.~\ref{fig:heuristic idea} depicts the outcome of CV calculating in stretched horizon dS holography from~\cite{Susskind:2021esx,Chapman:2021eyy,Galante:2022nhj,Jorstad:2022mls}. Similar considerations apply to CA and CV2.0. What one sees is that extremal volume slices cease to exist for large or small enough $t$ on the stretched horizon. This occurs because as a result of extremization the outermost CV carriers approach and touch future or past infinity. In $d = 1$ this implies a singular derivative of the complexity with respect to $t$ and in $d \geq 2$ this implies on top of a divergence of complexity itself, which is the precise statement of the hyperfast growth.

In dS or SdS geometry, there are infinitely many other spatial slices that do not exhibit hyperfast growth. For example, constant global time slices of dS depicted with orange in Fig.~\ref{fig:heuristic idea} exhibit persistent exponential growth at late times. Of course, at this level, such slices are not covariantly defined and it is not clear if their volumes arise from a particular CAny proposal.

The key idea in the present paper is to find a family of codimension one objects that avoid the future infinity (to start with, and later also the past infinity) in a similar manner as orange slices do in Fig.~\ref{fig:heuristic idea}, which fall into the class of CAny proposals. 

As it turns out, we do not have to search far: CMC slices that appeared earlier in the context of holographic complexity in~\cite{Chandra:2022pgl,Belin:2022xmt,Jorstad:2023kmq} will have precisely the desired property. Such slices will bend towards the past or future light cone as their curvature, respectively, increases or decreases. Then, there should exist a class of holographic complexity notions that without fine-tuning avoids the hyperfast growth associated with touching $\mathcal{I}^+$ or $\mathcal{I}^-$, or at best both, rendering the observables finite during the time evolution.

\section{The relevant class of complexity proposals}

CAny proposals~\cite{Belin:2021bga,Belin:2022xmt} are defined in a two-step procedure. First one defines a boundary (here: stretched horizon) anchored geometric region using extremization and, subsequently, one characterizes it in terms of, in general, another geometric functional yielding a non-negative number - a value of holographic complexity. Of course, the challenge lies in carving out the space of such functionals, which gives rise to the linear growth and the switchback effect for AdS black holes. What is known so far are several classes of objects specified by continuous parameters for which these properties have been demonstrated.

In our work, we will be interested in (spatial) volumes of stretched horizon-anchored CMC slices. Maximal volume slices giving rise to CV fall into this class, but we will be clearly interested in other members. Along the lines of CAny, they can be obtained by extremizing
\begin{equation}\label{eq:regions CMC}
\begin{aligned}
    \mathcal{C}_{\rm CMC}=\tfrac{1}{G_N}\biggl[&\alpha_+\int_{\Sigma_+}\rmd^d\sigma_+\,\sqrt{h}+\alpha_-\int_{\Sigma_-}\rmd^d\sigma_{-}\,\sqrt{h}\\
    &+\alpha_B\int_{\mathcal{M}}\rmd^{d+1}x\sqrt{-g}\biggr],
\end{aligned}
\end{equation}
where $\mathcal{M}$ is the codimension-zero bulk region that in the end will play no role; $\Sigma_\pm$ are its, crucial for us, future and past boundaries with general coordinates $\sigma_{\pm}$ and $\alpha_\pm$, $\alpha_B$ are positive constants. 

The extremization of~\eqref{eq:regions CMC} confirms $\Sigma_\pm$ are CMC slices
\begin{equation}
    \eval{{K}}_{\Sigma_\epsilon} = -\epsilon\frac{\alpha_B}{\alpha_\epsilon},\quad\epsilon=\pm~,
\end{equation}
where $K$ is the trace of the extrinsic curvature with normal vectors to both $\Sigma_{\pm}$ chosen to be future-oriented. 

Our CAny complexity carrier will be 
\begin{equation}\label{eq:Volepsilon}
    \mathcal{C}^\epsilon \equiv \frac{1}{G_N}\int_{\Sigma_\epsilon}\rmd^d\sigma_{\epsilon}\,\sqrt{h} \,,
\end{equation}
where we are free to pick either $\Sigma_{+}$ or $\Sigma_{-}$.
The results of~\cite{Belin:2022xmt} guarantee that~\eqref{eq:Volepsilon} is a valid CAny proposal.

\section{Late time growth}
The evaluation of the volume (\ref{eq:Volepsilon}) of the CMC slice $\Sigma_\epsilon$ can be recast as \cite{Jorstad:2023kmq},
\begin{equation}\label{eq:Anything complexity cod1}
\mathcal{C}^\epsilon=\frac{2\Omega_{d-1}}{G_N} \int_{{r}_{\rm st}}^{{r}_t} \frac{ {r}^{2(d-1)}\,\rmd {r}}{\sqrt{-\mathcal{U}(P_v^{\epsilon},\,{r})}}~,
\end{equation}
where $P_v^{\epsilon}$ is the conserved momentum in an analog particle motion problem,
\begin{equation}
\label{eq:potential arbitrary}
    \mathcal{U}(P_v^{\epsilon},\,{r}) = -f({r}) {r}^{2(d-1)} - \left(P_v^\epsilon - \lvert {K} \rvert \frac{\epsilon}{d}  {r}^d\right)^2
\end{equation}
is the particle's effective potential, whereas ${r}={r}_t$ is the turning point. The latter is the location where $\mathcal{U}(P_v^{\epsilon},\,{r}_t)=0$ or in geometric terms it is the tip of CMC ($r'(v) = 0$ there). We are interested in the time evolution of (\ref{eq:Volepsilon}) measured with respect to ${r}_{\rm st}$. Using the technology of~\cite{Belin:2021bga,Belin:2022xmt} one finds at late times
\begin{equation}\label{eq:Vol late times}
    \lim_{{t}\rightarrow\infty}\dv{{t}} \mathcal{C}^\epsilon =\frac{\Omega_{d-1}}{G_N} \sqrt{-f({r}_f) {r}_f^{2(d-1)}}
\end{equation}
where we consider solutions characterized by
\begin{equation}
\label{eq.accP}
\lim_{{t}\rightarrow \infty}\dv{P_v^\epsilon}{{t}}=0
\end{equation}
and ${r}_f\equiv\lim_{{t}\rightarrow \infty}{r}_t$ is the final value of the turning point. Notice that (\ref{eq:Vol late times}) does not depend on the particular value of $r_{\rm st}$. Condition~\eqref{eq.accP} can also be reformulated as finding the maximum of the potential (\ref{eq:potential arbitrary}): 
 \begin{equation}\label{eq:extr potential}
    \eval{\mathcal{U}}_{{r}_f}=0,\quad \eval{\partial_{{r}}\mathcal{U}}_{{r}_f}=0,\quad \eval{\partial^2_{{r}}\mathcal{U}}_{{r}_f}\leq0~.
 \end{equation}
We may define a function:
\begin{equation}\label{eq:rt location}
\begin{aligned}
   H(r,~K)=&4 {r}
   f\left({r}\right)
   \left((d-1)
   f'\left({r}\right)+{K}^
   2 {r}\right)\\
   &+4
   (d-1)^2
   f\left({r}\right){}^2
   +{r}^2
   f'\left({r}\right){}^
   2~,
\end{aligned}
\end{equation}
where the relations (\ref{eq:extr potential}) imply that late time growth of (\ref{eq:potential arbitrary}) with (\ref{eq:regions CMC}) is achieved when one can find the roots of
\begin{equation}
    H(r_f,~K)=0~,
\end{equation}
for some choice of $K$. We now specialize in asymptotically dS backgrounds, employing the factor (\ref{eq:blackening factor}). We discuss different cases under our proposal.
\begin{itemize}
    \item \textbf{Empty dS space}, $\mu=0$,
    \begin{equation}\label{eq:rf dS}
    {r}_f^2 = \frac{{K}^2-2 d(d-1)\pm|{K}| \sqrt{{K}^2-4(d-1)}}{2( {K}^2- d^2)}~.
\end{equation}
Then, in order to have at least one turning point at late times, ${r}_f\in\mathbb{R}$, in empty dS space with $d\geq2$ spatial dimensions, we find:
\begin{equation}\label{eq:main result}
    \abs{{K}} \geq {K}_{\rm crit,\, dS}= 2\sqrt{d-1}~.
\end{equation}
The CMC slices obeying this bound are displayed in Fig. \ref{fig:geodesics}.
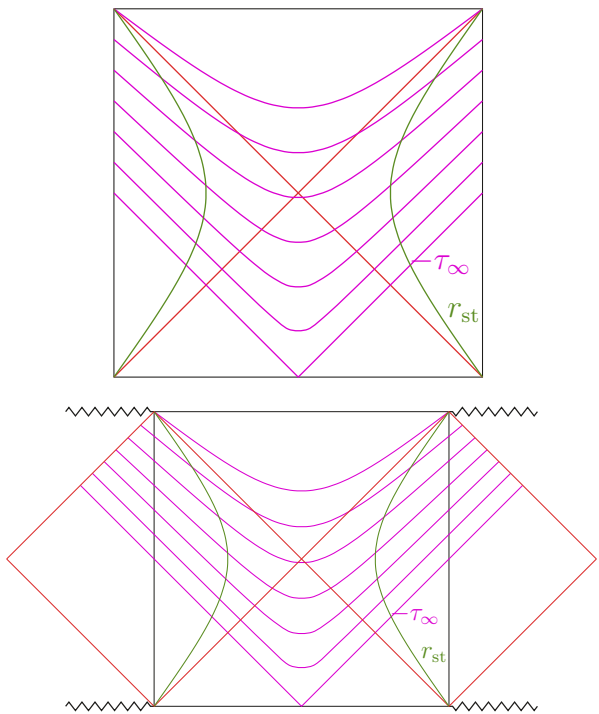
\begin{figure}[t!]
\centering
    \begin{tikzpicture}[scale=2.45]
        \draw[frame] (-1,-1)--(-1,1)--(1,1)--(1,-1)--cycle;
        \draw[uv] (-1,-1) -- (1,1);
        \draw[uv] (1,-1) -- (-1,1);
        \draw[lijn] (-1,0)--(0,-1)--(1,0);
        \node[lijn] at (0.77,-0.377) {\large $-\tau_\infty$};
        \def\Nlines{6}
        \foreach \i [evaluate={\c=(\Nlines+\i)/(2*\Nlines+1); \cs=sin(90*\c);}] in {1,...,\Nlines}{
        \draw[lijn, samples=\Nsamples, smooth, variable=\x, domain=0:2]
            plot(1-\x,{(\Nlines-\i+1)/\Nlines-kruskal(\x*pi/4,\cs))});
        }
        \node[r, anchor=south] at (0.9, -0.75) {\large ${r}_\text{st}$};
        \draw[r, samples=\Nsamples, smooth, variable=\y, domain=0:2]
                plot({1-kruskal(\y*pi/4,sin(45))},\y-1);
        \draw[r, samples=\Nsamples, smooth, variable=\y, domain=0:2]
                plot({-1+kruskal(\y*pi/4,sin(45))},\y-1);
    \end{tikzpicture}
     \\ \vspace{1em}
    \scalebox{0.8}{
    \begin{tikzpicture}[scale=2.45]
        \draw[frame] (-1,-1)--(-1,1)--(1,1)--(1,-1)--cycle;
        \draw[lijn] (-1.5,0.5)--(0,-1)--(1.5,0.5);
        \node[lijn] at (0.77,-0.377) {\large $-\tau_\infty$};
        \def\Nlines{6}
        \foreach \i [evaluate={\c=(\Nlines+\i)/(2*\Nlines+1); \cs=sin(90*\c);}] in {1,...,\Nlines}{
        \draw[lijn, samples=\Nsamples, smooth, variable=\x, domain={-0.1*(\i-1)}:{2+0.1*(\i-1)}]
            plot(1-\x,{(\Nlines-\i+1)/\Nlines-kruskal(\x*pi/4,\cs))});
        }
        \fill[white] (1,1)--(2,0)--(2,1);
        \fill[white] (-1,1)--(-2,0)--(-2,1);
        \draw[frame, singularity] (-1.6,1) -- (-1,1);
        \draw[frame, singularity] (-1.6,-1) -- (-1,-1);
        \draw[frame, singularity] (1.6,1) -- (1,1);
        \draw[frame, singularity] (1.6,-1) -- (1,-1);
        \draw[uv] (-2,0)--(-1,-1)--(1,1)--(2,0);
        \draw[uv] (2,0)--(1,-1)--(-1,1)--(-2,0);
        \node[r, anchor=south] at (0.9, -0.75) {\large ${r}_\text{st}$};
        \draw[r, samples=\Nsamples, smooth, variable=\y, domain=0:2]
                plot({1-kruskal(\y*pi/4,sin(45))},\y-1);
        \draw[r, samples=\Nsamples, smooth, variable=\y, domain=0:2]
                plot({-1+kruskal(\y*pi/4,sin(45))},\y-1);
    \end{tikzpicture}
    }
    \caption{CMC slices for ${K}\geq {K}_{\rm crit,\, dS}$ in empty dS$_{d+1}$ space (above) and SdS$_{d+1}$ (below). All the slices remain bounded below $\mathcal{I}^+$ and the corresponding complexity observable (\ref{eq:Anything complexity cod1}) generically displays a late-time linear growth (\ref{eq:Vol late times}), except for some fine-tuned situations discussed in the main text. The solutions with ${K}<0$ can be obtained by a top-bottom reflection.}
    \label{fig:geodesics}
\end{figure}
However, notice that the relation (\ref{eq:main result}) is not valid when $d=1$, since for $d=1, \abs{{K}} < 1$ equation (\ref{eq:rf dS}) does not have a valid solution; instead one finds ${K}\geq 1$ for the CMC slices to evolve at arbitrarily late times.

\item For the \textbf{Nariai black hole spacetime}, $\mu=\mu_N$, one finds that $f({r}_f)=f'({r}_f)=0$ at the location 
\begin{equation}\label{eq:rf N}
{r}_f=\sqrt{\tfrac{d-2}{d}}~,    
\end{equation}
such that the turning point coincides with the cosmological horizon. However, for ${r}_f$ to be the final slice, it also needs to be a maximum of the potential $\mathcal{U}(P^\epsilon_v,\,{r})$ in (\ref{eq:potential arbitrary}), which leads to,
\begin{equation}\label{eq:Kcrit N}
    \abs{{K}}\geq{K}_{\rm crit,\, N}\equiv \sqrt{d}~.
\end{equation}

\item For generic $\mu$, one cannot derive closed-form solutions for ${r}_f$ in (\ref{eq:rt location}), except for the SdS$_3$ space, which is locally identical to dS$_3$. We explicitly find that (\ref{eq:main result}) is always respected in such a case. For higher dimensions and generic $\mu$, the bounds on $\abs{{K}}$ will lay between (\ref{eq:main result}) and (\ref{eq:Kcrit N}) \cite{Aguilar-Gutierrez:2023pnn}. Such black holes are unstable and decay in empty dS$_{d+1}$ space \cite{Ginsparg:1982rs}. The analysis for  generic Reissner-Nordstrom-de Sitter (RNdS) black holes is shown in the supplementary material.
\end{itemize}

Note that the solutions for (\ref{eq:rf dS}) in $d=1$ and $d=2$ as well as for SdS$_3$, lead to ${r}_f\rightarrow\infty$ when $\abs{{K}}={K}_{\rm crit}$.
We find exponential growth for (\ref{eq:Anything complexity cod1}) in these two very special cases. For $d = 3$ and higher we find finite $r_{f}$ for $K = K_{\rm{crit}}$, which translates to the linear growth. 

\section{Restoring time symmetry}
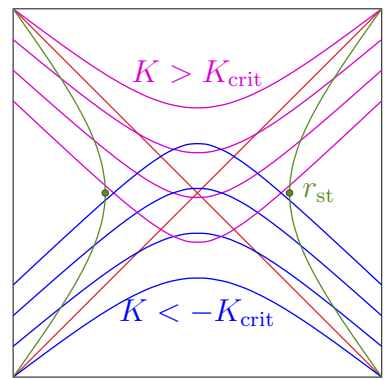
\begin{figure}[t!]
\centering
    \begin{tikzpicture}[scale=2.45]
        \draw[frame] (-1,-1)--(-1,1)--(1,1)--(1,-1)--cycle;
        \draw[uv] (-1,-1) -- (1,1);
        \draw[uv] (1,-1) -- (-1,1);
        \def\Nlines{6}
        \foreach \i [evaluate={\c=(\Nlines+\i)/(2*\Nlines+1); \cs=sin(90*\c);}] in {1,...,4}{
        \draw[lijn, samples=\Nsamples, smooth, variable=\x, domain=0:2]
            plot(1-\x,{(\Nlines-\i+1)/\Nlines-kruskal(\x*pi/4,\cs))});
        \draw[lijn,blue, samples=\Nsamples, smooth, variable=\x, domain=0:2]
            plot(1-\x,{-(\Nlines-\i+1)/\Nlines+kruskal(\x*pi/4,\cs))});
        }
        \node[r, anchor=east] at (0.8, 0) {\large ${r}_\text{st}$};
        \draw[r, samples=\Nsamples, smooth, variable=\y, domain=0:2]
                plot({1-kruskal(\y*pi/4,sin(45))},\y-1);
        \draw[r, samples=\Nsamples, smooth, variable=\y, domain=0:2]
                plot({-1+kruskal(\y*pi/4,sin(45))},\y-1);
        \node[lijn,blue] at (0, -0.65) {\large ${K} < -K_{\rm crit}$};
        \node[lijn] at (0, 0.65) {\large ${K} > K_{\rm crit}$};
        \draw[lijn,zwart!50!groen,line width=0.3,fill=groen] (0.5, 0) circle (0.5pt);
        \draw[lijn,zwart!50!groen,line width=0.3,fill=groen] (-0.5, 0) circle (0.5pt);
    \end{tikzpicture}
    \caption{Our time symmetric complexity proposal~\eqref{eq.Csym} in empty ${\rm dS}_{d+1}$ allowing both early- and late-time linear growth. At negative times, the CMC with ${K} < -K_{\rm crit}$ dominates, shown in blue; while at positive times, the CMC with ${K} > K_{\rm crit}$ dominates. The exchange of dominance at $t=0$ between CMC slices is indicated by the green dots on the stretched horizon.}
    \label{fig:timesymmetric}
\end{figure}
Although the rate of growth of the observables in (\ref{eq:Volepsilon}) evaluated on the CMC slices 
asymptotes to a constant value at late times (\ref{eq:Vol late times}) when ${K}>K_{\rm{crit}}$ 
the CMC slices still hit $\mathcal{I}^-$ at minus the critical time, 
as illustrated in Fig. \ref{fig:geodesics}. This produces hyperfast behavior in the past. The opposite case occurs by symmetry for ${K}<-K_{\rm{crit}}$.

A natural way to restore time-reversal symmetry in the observables is to modify the second step of the CAny prescription so that it selects the result with the minimal value among the slices with a given value for $\abs{{K}}$, 
\begin{equation}
\label{eq.Csym}
    \mathcal{C}_{\rm sym} = \min_{\epsilon = +, -} \mathcal{C}^\epsilon \,.
\end{equation}
The minimization is performed over the existing slices, so technically it is only a minor modification. This procedure does not alter the conclusion that the constructed covariant notions are complexity proposals, as the linear growth and the switchback effect for AdS black holes remain present. We show this explicitly in the \textit{Appendix}.\footnote{One of the authors has recently shown that this class of proposals also satisfies the switchback effect in SdS space \cite{Aguilar-Gutierrez:2023pnn}.} As a result, this idea can be thought of as a further enlargement of the space of CAny proposals and might even be advantageous when considering more complicated black holes in AdS.

In the dS case, our improved proposal~\eqref{eq.Csym} will receive a contribution from a slice with ${K}<0$ at early times, and ${K}>0$ at late times, as shown in Fig.~\ref{fig:timesymmetric}. A potential subtlety with this generalization is that the complexity growth rate (\ref{eq:Vol late times}) might become discontinuous at the time when the change of CMC slice occurs. However, this is in principle allowed in the definition of holographic complexity proposals \cite{Belin:2021bga,Belin:2022xmt}. Indeed, the CA proposal applied to a three dimensional AdS black hole generically exhibits the same
kind of behavior~\cite{Carmi:2017jqz}.

\section{Discussion}
Our paper demonstrates that the hyperfast growth of holographic complexity in asymptotically dS spacetimes, as found earlier in the CV, CA, and CV2.0 proposals, is not a universal feature in the CAny landscape. Employing volumes of codimension-one CMC slices, being members of CAny family, we show that holographic complexity can exhibit persistent linear or exponential growth in asymptotically dS universes. Physically, this exponential behavior occurs when the final slice asymptotes the future/past infinity of the inflating region. While a linear growth can be also obtained upon cutting out the dS geometry past some late time slice~\cite{Jorstad:2022mls}, we obtained it without modifying dS geometry in any way.

From the perspective of dS holography, it is tempting to speculate that the presence or the absence of hyperfast growth is related to the choice of a penalty schedule in the microscopic definition of complexity, i.e. different designations which operations are hard and which are easy to implement. It would be interesting to study it, as well as the protocol in (\ref{eq.Csym}), in either a class SYK models associated with JT gravity with positive cosmological constant~\cite{Susskind:2022bia,Rahman:2022jsf,Goel:2023svz,Blommaert:2023opb} or in quantum circuit toy models of de Sitter space~\cite{Bao:2017iye,Bao:2017qmt,Niermann:2021wco,Cao:2023gkw}.

More along these lines, specializing in CV proposal in two spacetime dimensions, where the complexity carriers are geodesics, it is known that in dS past the critical time on a stretched horizon, there are no spatial geodesics. However, using a closed-form expression for a geodesic distance one obtains an answer with both real and imaginary parts~\cite{Chapman:2022mqd,Aalsma:2022eru}. While one may speculate about non-orthodox interpretation in terms of complexity, e.g. with real part accounting for unitary and imaginary part for possible non-unitary gates, we find it important to stress that our final result~\eqref{eq.Csym} does not require any departure from a standard counting interpretation of unitary gates.

Furthermore, the space of CAny proposals is vast, and arguably one of the main open problems for the field of holographic complexity is to study it in a more systematic manner. To this end, our results show the existence of a so far unrecognized structure in the CAny landscape stemming from the presence (so far demonstrated for CV, CA, and CV2.0) or the absence of the hyperfast growth demonstrated here for CMC complexity carriers. One intriguing future research direction would be to find more members of the hyperfast growth escaping CAny proposals and, another, to seek other structures present. On the former front, we want to emphasize that there is a continuum of CAny proposals that do not exhibit hyperfast growth, as encapsulated by~\eqref{eq:main result}.

We also want to highlight a potentially puzzling feature for a class of CAny proposals we considered, which to the best of our knowledge has not been previously seen in the literature. As illustrated in Fig.~\ref{fig:geodesics}, asymmetric time evolution may occur such that hyperfast growth is observed in the past or future, while the linear or exponential growth remains for the late or early time regime respectively. If we want to assign a Nielsen unitary complexity~\cite{Nielsen,Chapman:2021jbh} interpretation to this setting, then the complexity of a unitary is the same as its inverse. This implies time asymmetric quantities in time-symmetric setups either do not capture (this type of) complexity or the considered time evolution is not unitary. The same occurs for CAny proposals on CMC slices even for AdS planar black holes with the location of the early/late turning point not being time-reflection symmetric.

The time symmetry in the observables can be restored by introducing a covariant protocol that alternates between CMC slices of opposing sign, where the slice that minimizes complexity is chosen. This consideration led us to a new CAny proposal encapsulated by~\eqref{eq.Csym}, which is time-symmetric. Notice, however, that we could have chosen instead a protocol maximizing complexity over CMC slices, or even averaging, instead of doing the minimization that we proposed. However, in such cases, the hyperfast growth would not be avoided anymore.

Finally, let us reiterate that the defining features for CAny proposals are the late-time linear growth and the switchback effect for AdS black holes. If one were to add to this list the hyperfast growth in dS, our paper could be then viewed as ruling out a subclass of CAny proposals.

\vspace{20 pt}

\begin{acknowledgments}
\noindent \emph{Acknowledgments:} We would like to thank Stefano Baiguera, Shira Chapman, Dami\'{a}n Galante, Eivind J{\o}rstad, Ayan K. Patra, Juan F. Pedraza, Leonard Susskind, Qi-Feng Wu, and Nicolò Zenoni for useful discussions on de Sitter space and complexity, and Alexandre Serantes for a collaboration on a related topic. SEAG thanks the University of Amsterdam and the Delta Institute for Theoretical Physics for their hospitality and support during the research. The work of SEAG is partially supported by the FWO Research Project G0H9318N and the inter-university project iBOF/21/084.
\end{acknowledgments}

\appendix
\section{Late time growth in Reissner-Nordstrom-de Sitter space}
We consider a generalization of the previous analysis for electrically or magnetically charged black holes in asymptotically dS space, known as the Reissner-Nordstrom-de Sitter (RNdS) space. The blackening factor in $(d+1)-$dimensions in (\ref{eq:blackening factor}) is modified to
\begin{equation}\label{eq:blackening factor RNdS}
    f(r)=1-r^2-\tfrac{2\mu}{r^{d-2}}+\tfrac{q^2}{r^{2(d-2)}}
\end{equation}
where, $q$ is a parameter related to the electric or magnetic charge of the black hole, which we will consider $q>0$ through the discussion. The three positive roots of (\ref{eq:blackening factor RNdS}) for $d>2$ represent the inner and outer black hole horizon, as well as the cosmological horizon; however, the real solutions are only present when there are bounds on the parameters $\mu$ and $q$.
There exists a black hole with maximal mass and charge parameters, $\mu=\mu_U$ and $q=q_U$ respectively, denoted as the ultracold (``U") solution, for which \cite{Morvan:2022aon}
\begin{equation}
    \mu_U=\tfrac{2}{d}\qty(\tfrac{(d-2)^2}{d(d-1)})^{\tfrac{d-2}{2}}~,\quad q_U=\tfrac{1}{\sqrt{d-1}}\qty(\tfrac{(d-2)^2}{d(d-1)})^{\tfrac{d-2}{2}}~.
\end{equation}
In these conditions, the outer, inner, and cosmological horizons have the same radius, $r_U$, for which
\begin{equation}
    f(r_U)=f'(r_U)=f''(r_U)=0,\quad r_U=\tfrac{d-2}{\sqrt{d(d-1)}}~,
\end{equation}
which indicates that $r_f=r_U$ is a root for (\ref{eq:rt location}). Moreover, there are no bounds for $\abs{K}$ resulting from the conditions (\ref{eq:extr potential}) in this limit.

We will evaluate $H(r,\,K)$ in (\ref{eq:rt location}) with the roots in (\ref{eq:rf dS}, \ref{eq:rf N}) while keeping the mass and charge of the black hole arbitrary. Defining $m\equiv \mu/\mu_U$ and $\rho\equiv q/q_U$, one might express:
\begin{widetext}
\begin{align}
    &\begin{aligned}
        H\qty({r}^{\rm (dS)}_f,\,{K}_{\rm crit,\, dS})=4 \Bigl(&4 (d-2)^{3 (d-2)} d^{2-d} (d-1)^{4-2 d} \left(m^2+\rho ^2\right)-4
   (d-2)^{\frac{9 (d-2)}{2}} d^{3-\frac{3 d}{2}} (d-1)^{5-3 d} m \rho ^2\\
   &-8(d-2)^{\frac{3 (d-2)}{2}} d^{-d/2} (d-1)^{3-d} m+(d-2)^{6 (d-2)} d^{4-2 d}(d-1)^{6-4 d} \rho ^4\Bigr)~,
    \end{aligned}\label{eq:HdS}\\
    &\begin{aligned}
    H\qty({r}^{\rm (N)}_f,\,K)=\tfrac{4}{(d-1)^2 d^2} \Bigl(&4 d^4 (m-1)^2+d^3 \left(K^2 \left(-4 m+\rho ^2+3\right)-4 (m-1)\left(2 m+\rho ^2-3\right)\right)+16 K^2 (m-1)\\
    &+d^2 \left(4 K^2 \left(5 m-\rho^2-4\right)+\left(2 m+\rho ^2-3\right)^2\right)+4 d K^2 \left(-8 m+\rho^2+7\right)\Bigr)~.
    \end{aligned}\label{eq:HN}
\end{align}
\end{widetext}
We deduce that (\ref{eq:HdS}) is negative for all $d\geq3$ for the allowed parameter space of $m,~q\in(0,\,1)$, while (\ref{eq:HN}) is positive. Moreover, $H\qty({r}^{\rm (dS)}_f,\,K)$ becomes more negative as we increase $\abs{K}>{K}_{\rm crit,\, dS}$. Then, according to the \emph{intermediate value theorem}, there will exist at least a real root ${r}_f\in \qty[{r}^{\rm (dS)}_f,\,{r}_{\rm U}]$ for general RNdS$_{d+1}$ space.

\section{Switchback effect in asymptotically AdS planar black holes}
In this appendix, we show that the CAny proposals in (\ref{eq:regions CMC}, \ref{eq:Volepsilon}) reproduce the switchback effect in AdS planar black holes. The late-time linear growth property for the proposals under consideration has been studied in \cite{Belin:2022xmt,Jorstad:2023kmq}.

The asymptotically AdS$_{d+1}$ planar black hole metric can be expressed 
\begin{equation}
\begin{aligned}\label{eq:blackening factor AdS BH}
    \rmd s^2&=-f(r)\rmd t^2+\tfrac{\rmd r^2}{f(r)}+r^2\rmd \vec{x}^2~,\\
    f(r)&=r^2\qty(1-\tfrac{r_h^d}{r^d})~;
\end{aligned}
\end{equation}
where $r_h$ is the location of the black hole horizon, and $\vec{x}$ a ($d-1$)-dimensional vector.

\begin{figure*}
    \centering
    {\includegraphics[width=0.33\textwidth]{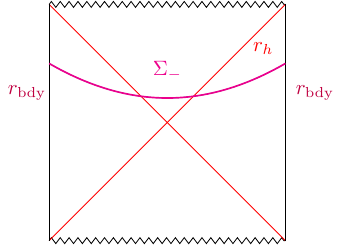}}\hfill{\includegraphics[width=0.33\textwidth]{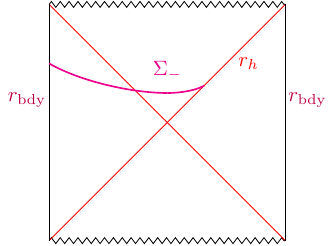}}\hfill{\includegraphics[width=0.33\textwidth]{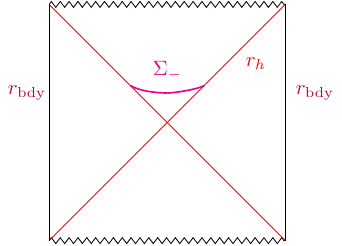}}
    \caption{Extremal complexity surface $\Sigma_\epsilon$ (pink) in (\ref{eq:total SW complexity}) for $\epsilon=-$ in a planar AdS black hole. \textit{Left}: $\mathcal{C}^-(t_L,\,t_R)$; \textit{center}: $\mathcal{C}^-(V_R,~t_L)$; and \textit{right}: $\mathcal{C}^-(V_R,~U_L)$. $r_{\rm bdy}$ (purple) denotes the radial cutoff of the asymptotic boundary, and $r_h$ (red) the event horizon.}
    \label{fig:anchoringAdS}
\end{figure*}

We will consider a coordinate transformation from the EF to Kruskal coordinates, which we define by
\begin{equation}\label{eq:Kruskal coord}
    U=\rme^{-\tfrac{f'(r_h)}{2}u}~,\quad V=-\rme^{\tfrac{f'(r_h)}{2}v}~.
\end{equation}
The geometry for shockwave perturbations sent along $U=0$ as \cite{Belin:2021bga,Belin:2022xmt} can be then described as
\begin{align}\label{eq:metric SW alpha}
        \rmd s^2=&-2A\qty(U[V+\alpha_i\Theta(U)])
    \rmd U\rmd V\\
    &+B\qty(U[V+\alpha_i\Theta(U)])\rmd \vec{x}^2~,
\end{align}
where
\begin{equation}
\begin{aligned}
    A\qty(UV)&\equiv-\tfrac{2}{UV}\tfrac{f(r)}{f'(r_h)^2}~,\quad B\qty(UV)\equiv r^2~,\\ \alpha_i&=2\rme^{-\tfrac{f'(r_h)}{2}(t_*^{(\rm b)}\pm t_i)}~.\label{eq:factors a b}
\end{aligned}
\end{equation}
Here, $t_i$, with $i\in1,\dots n$, are the shockwave insertion times with respect to the asymptotic boundary, where we will consider the total number of shockwaves $n$ to be even; the $\pm$ sign indicates the direction that the shockwaves are sent to; and $t_*^{(\rm b)}$ is the scrambling time of the AdS black hole. We are interested in an alternating order for the insertion times, i.e.
\begin{equation}
t_{2k+1}>t_{2k},\quad t_{2k}< t_{2k-1}~,
\end{equation}
where $k\in1,\dots, n/2$. Moreover, we consider the shockwave regime where $\abs{t_{i+1}-t_i}\gg t_*^{(\rm b)}$. Under these conditions, one may express the complexity in the alternating shockwave as \cite{Belin:2021bga,Belin:2022xmt}:
\begin{equation}\label{eq:total SW complexity}
\begin{aligned}
    &{\mathcal{C}}^\epsilon(t_L,\,t_R)=\mathcal{C}^\epsilon(t_R,\,V_1)+\mathcal{C}^\epsilon(V_1+\alpha,\,U_2)+\dots\\
    &+\mathcal{C}^\epsilon(U_{n-1}-\alpha_{n-1},\,V_n)+\mathcal{C}^\epsilon(V_{n-1}+\alpha_{n-1},\,t_L)~,
\end{aligned}
\end{equation}
where $\mathcal{C}^\epsilon(\cdot,\,\cdot)$ denotes the contributions from $\Sigma_\epsilon$ with two fixed endpoints and all endpoints are located either on the left/right event horizon ($r_h$) or asymptotic infinity. The different cases are illustrated in Fig. \ref{fig:anchoringAdS}.\footnote{In the following, we will work in scaled coordinate where the AdS scale $\ell_{\rm AdS}=1$.}

The different contributions in (\ref{eq:metric SW alpha}) can be expressed with EF coordinates (\ref{eq:starting metric}) and considering symmetric time evolution $t_L=t_R=t/2$ as:
\begin{widetext}
\begin{align}
    &\mathcal{C}^\epsilon(t_R,\,V_L)=\mathcal{C}^\epsilon(V_R,\,t_L)=-\frac{\mathcal{V}_x}{G_N}a(r_t)\sqrt{-f(r_t)r_t^{2(d-1)}}\qty(\int^{r_{\rm bdy}}_{r_t}+\int^{r_h}_{r_t})\frac{\left(P_v^\epsilon + \frac{\epsilon L\abs{K}}{d} r^{d} \right)\rmd r}{f(r)\sqrt{-\mathcal{U}(P_v^\epsilon, r)}}~,\nonumber\\
    &\mathcal{C}^\epsilon(V_R,\,U_L)=-\frac{2\mathcal{V}_x}{G_N}a(r_t)\sqrt{-f(r_t)r_t^{2(d-1)}}\int^{r_h}_{r_t}\frac{\left(P_v^\epsilon + \frac{\epsilon L\abs{K}}{d} r^{d} \right)\rmd r}{f(r)\sqrt{-\mathcal{U}(P_v^\epsilon, r)}}~.\label{eq:Cepsilon VRUL}
\end{align}
\end{widetext}
where $r_{\rm bdy}$ represents a cutoff radial location of the asymptotic boundary. One can perform a very similar analysis of the time dependence of the CAny proposals in (\ref{eq:regions CMC}, \ref{eq:Volepsilon}) to the one in the main text. One of the differences, however, is that the effective potential $\mathcal{U}(P_v^\epsilon,\,r)$ in (\ref{eq:potential arbitrary}) has a modification $K\rightarrow-K$ \cite{Belin:2022xmt}. Denoting again $\eval{K}_{\Sigma_+}=-\abs{K}$ and $\eval{K}_{\Sigma_-}=\abs{K}$, we have
\begin{equation}
    \mathcal{U}(P_v^{\epsilon},\,r)\equiv -f(r) r^{2(d-1)} - \left(P_v^\epsilon +\epsilon \frac{\abs{K}}{d}  r^d\right)^2~.
\end{equation}
To study the complexity growth evolution in the perturbed geometry, we must also evaluate its dependence on the location $u_{R,\,L}$, $v_{R,\,L}$ where $\Sigma_\epsilon$ intersects with the left/right horizon $r_h$. In this case, we derive
\begin{equation}
    v_R-v_t=\int_{r_t}^{r_h}\tfrac{\rmd r}{f(r)}\qty(1-\frac{P_v^\epsilon+\frac{\epsilon \abs{K}}{d}r^d}{\sqrt{-\mathcal{U}(P_v\epsilon,\,r)}})~,
\end{equation}
where $v_t=v_R(r_t)$. 

We may also perform the expansion around the final slice where (\ref{eq:extr potential}) allows us to approximate
\begin{equation}\label{eq: Approx Potential}
\lim_{r\rightarrow r_f}U(P_v^\epsilon,\,r)\simeq \tfrac{1}{2}(r-r_f)^2\mathcal{U}''(P_v^\epsilon,\,r)+\mathcal{O}(\abs{r-r_f}^3)~.
\end{equation}
The CAny proposal near the final turning point $r_f$ then can be evaluated as follows
\begin{equation}
\begin{aligned}
{\mathcal{C}^\epsilon}(V_R,\,U_L)=\tfrac{\mathcal{V}_{x}}{G_N}P_\infty^\epsilon v~,\quad P_\infty^\epsilon=\sqrt{-f(r_f)r_f^{2(d-1)}}
\end{aligned}
\end{equation}
where $r_f$ is a root of the function in (\ref{eq:rt location}).

The result above can be used to evaluate the contributions in (\ref{eq:total SW complexity}) as:
\begin{align}
{\mathcal{C}^\epsilon}(V_R,\,t_L)&=\frac{\mathcal{V}_x}{G_N}P^\epsilon_{\infty}\log{\rme^{t_L}V_R}~,\\
{\mathcal{C}^\epsilon}(V_R,\,U_L)&=\frac{\mathcal{V}_x}{G_N}P_{\infty}^\epsilon\log{U_LV_R}~,\\
{\mathcal{C}^\epsilon}(t_R,\,V_L)&=\frac{\mathcal{V}_x}{G_N}P_{\infty}^\epsilon\log{V_L\rme^{t_R}}~.
\end{align}
However, there will also be an early time contribution in the shockwave geometry, given by the term
\begin{equation}
{\mathcal{C}^\epsilon}(V_L,\,U_R)=\frac{\mathcal{V}_x}{G_N}P_{-\infty}^\epsilon\log{U_LV_R}~,
\end{equation}
where
\begin{equation}
    P_{-\infty}^\epsilon=\lim_{t\rightarrow-\infty}\sqrt{-f(r_I)r_I^{2(d-1)}}~,
\end{equation}
and $r_I=\lim_{t\rightarrow-\infty}r_t$, for which there is a sign flip in $K\rightarrow-K$. Using the blackening factor (\ref{eq:blackening factor AdS BH}), we explicitly find that $P_{\infty}^\epsilon=P_{-\infty}^\epsilon$. (\ref{eq:total SW complexity}) then transforms into
\begin{equation}\label{eq:Cepsilon tot}
\begin{aligned}
{\mathcal{C}^\epsilon}(t_L,\,t_R)\simeq \tfrac{\mathcal{V}_x\,P_{\infty}^\epsilon}{G_N}\Bigl[& U_1\rme^{t_R}+\log(U_1-\alpha)V_2+\dots\\
&+\log((V_n+\alpha_n)\rme^{t_L})\Bigr]~.
\end{aligned}
\end{equation}
Extremizing (\ref{eq:Cepsilon tot}) with respect to an arbitrary interception point ($V_i$, $U_i$) in the multiple shockwave geometry,
\begin{equation}
\dv{{\mathcal{C}^\epsilon}(t_L,\,t_R)}{V_i}=0~,\quad \dv{{\mathcal{C}^\epsilon}(t_L,\,t_R)}{U_i}=0~,
\end{equation}
leads us to the location
\begin{align}
V_i=-\frac{\alpha_i}{2}~,\quad U_i=\frac{\alpha_i}{2}~.
\end{align}
Replacing the interception points into (\ref{eq:Cepsilon tot}) generates:
\begin{equation}\label{eq:Intermediate switchback}
{\mathcal{C}^\epsilon}\simeq\tfrac{\mathcal{V}_x\,P_{+\infty}^\epsilon}{G_N}\qty(t_R+t_L+2\qty(\sum_{k=1}^nt_k-nt_*^{(b)}))~,
\end{equation}
where the result is expressed up to the addition of constant terms. Given the ordering of the insertion times, we reproduce the switchback effect property \cite{Belin:2021bga,Belin:2022xmt},
\begin{equation}\label{eq:switchback result}
    \mathcal{C}^\epsilon\propto\abs{t_R+t_1}+\abs{t_2-t_1}+\dots+\abs{t_n-t_L}-2nt^{(b)}_*~.
\end{equation}
Notice that the minimization protocol introduced in (\ref{eq.Csym}) does not alter the result above.

\bibliographystyle{utphys}
\bibliography{references} 

\end{document}